\documentstyle [10pt] {article}
\input psfig.tex
\begin{document}
\begin{center}
\begin{bf}
\large\bf{Global Inflow and Outflow Solutions (GIOS) around a Black Hole}
\vskip0.35cm

%\vskip0.35cm
%\end{large}
\end{bf}
Sandip K. Chakrabarti

%\end{bf}
\vskip0.5cm

S. N. Bose National Centre For Basic Sciences,\\
JD Block, Salt Lake, Sector-III, Calcutta-700091\\ 	
e-mail: chakraba@boson.bose.res.in\\
\vskip0.5cm
\end{center}

\baselineskip = 12 true pt

\def\lsim{\lower.5ex\hbox{$\; \buildrel < \over \sim \;$}}
\def\gsim{\lower.5ex\hbox{$\; \buildrel > \over \sim \;$}}
	
\begin{abstract}
Twenty five years have passed by since models of accretions and jets have
separately emerged. Today, it is understood that these two objects are
related to each other in a fundamental way. In a binary system, matter from 
an accretion disk enters into a black hole.
A part of it is bounced back because of the centrifugal barrier, radiation pressure
or magnetohydrodynamic effects, to form jets and bipolar outflows which 
carry away excess angular momentum. In the case of AGNs containing black holes, 
accretion disks form out of stellar winds and similar processes as above form 
cosmic radio jets.  We present a general review of the study of the accretion disks 
and outflows in a coherent manner, especially emphasizing global inflow-outflow 
solutions (GIOS).  We also present a few observational consequences of wind 
production from the accretion disks on spectral properties of the accretion disks.

\end{abstract}

\noindent To appear in the proceedings of the Mini Workshop on Fluid Dynamics at the
Dacca University (August, 1998). Ed. A. Hosain.

\section{Introduction}

Study of accretion disks and outflows around black holes began twenty-five years ago [1-2].
The subject has evolved considerably since then and it is now clear that these two apparently
dissimilar objects are related to each other. The accretion 
solutions of purely rotating disk have been improved to include the effect of radiation pressure
[3-7] (See Chakrabarti [8] for a review.).
Jet solutions have changed from speculative ideas such as de-Laval nozzles [9]
to electrodynamically acceleration model [10], self-similar 
centrifugally driven outflows [11], `cauldrons' [12]  etc. Centrifugally driven outflows
are subsequently modified to include accretion disks [13].
Chakrabarti \& Bhaskaran [14] (see also, Contopoulos, [15])
showed that it is easier produce outflows from a sub-Keplerian inflow.

Parallelly, efforts were on to study the accretion and jets within the same framework. 
Chakrabarti [16-17] found that the same solution of purely rotating flows around a 
black hole could describe accretion flows on the equatorial plane and pre-jet matters 
near the axis. With a natural angular momentum distribution of $l(r) = c \lambda (r)^n$, 
(where $c$ and $n$ are constants and $\lambda$ is the von Zeipel parameter) 
it was found that for large $c$ and small $n$ ($n<1$), solutions are regular on the equatorial plane 
and they describe thick accretion disks. For small $c$ and large $n$ ($n>1$), the solutions
are regular on the axis and they describe pre-jet matters. It was speculated that
some viscous process might be responsible to change the parameters from one set to the
other. Even in the absence of viscosity, constant angular momentum flow was found to bounce back
from the centrifugal barrier in numerical simulations of accretion flows [18]. Eggum, Coroniti \& Katz [19]
considered radiatively outflows emerging from a Keplerian disk.

Further progress of this topic required a fundamental understanding  of these flows which 
emerged in the late 1980s. It was found out just as Bondi solution [20] of accretion 
onto stars is related to outflow solutions [21],
fundamentally transonic black hole accretions and winds are also related to each other. 
All possible accretion and wind type solutions are found, including solutions which
may contain standing shocks and the entire parameter space is classified
according to the nature of solutions [22-23].
Fig. 1 (taken from Chakrabarti, [24]) shows these solutions (Mach number is 
plotted against logarithmic radial distance [in Units of the Schwarzschild Radius]
and outer boxes, and specific energy is plotted against specific angular momentum 
in the central box) and the classification of the parameter space when the Kerr
parameter $a=0.5$ and when the equatorial plane solutions are considered (similar 
solutions are present for conical flows in winds as well, see [23]).
The inward pointing arrows indicate accretion solutions
and the outward pointing arrows indicate wind solutions. The flow 
from regions I, O pass through the inner sonic point  and outer sonic 
point respectively. Those from NSA have no shock in accretion, from 
SA have shocks in accretion, from NSW have no shocks in winds, and SW have shocks
in winds respectively.
The Global Inflow Outflow Solutions (GIOS) as will be described in \S 3 combine one solution of
each kind to produce wind out of accretion. The horizontal line in the central box 
denotes the rest mass of the inflow. Note that the outflows are produced only when the
specific energy is higher than the rest mass energy. Flow with with lesser energy
produces solutions with closed topologies (I* and O*). 
When viscosity was added the nature of the solutions changed fundamentally [25]
(see, [26-27] for details) allowing matter to directly come out of a Keplerian
disk and enter into a black hole through the sonic point.
These solutions, both for the accretion and winds have been verified by complete time-dependent
simulations [28-31]. In the case of steady
state solutions, outflows are found to occur between the centrifugal barrier and the
funnel wall (Fig. 2a below) while in a non-steady solution the outflow
could spread in regions outside the centrifugal barrier as well.

\section{Recent Progresses in Accretion Disk Solutions}

Fig. 2a shows the general picture that emerges of the non-magnetized inflow and outflow 
around a black hole. Centrifugal force tends to fight against gravity close to 
a black hole to produce a denser region called CENBOL (centrifugal barrier supported 
boundary layer). Matter farther out rotates in a Keplerian disk, but close to the black hole
the flow is puffed up due to radiation pressure or ion pressure depending on whether the
accretion rate is sufficiently high or not. The two-dimensional nature of the density
distribution is given in Chakrabarti [24].

Chakrabarti \& Titarchuk [32] using this generalized accretion solution suggested that
hard and soft states of a black hole could be understood simply by re-distribution of
matter between Keplerian and sub-Keplerian components which is effected by 
variation of viscosity in the flow in the vertical direction. This general conclusions is verified
observationally (Zhang et al. 1997).
The constancy of energy spectral index $\alpha_e$ ($F_\nu \propto \nu^{-\alpha_e}$)
separately in hard and soft states, as well as possible shifting of the inner edge
with the Keplerian component with accretion rate as predicted by the 
general accretion solution [26] have also been verified [34-36].
It was also recognized that the non-stationary solutions of the accretion flows  [37-38]
might cause quasi-periodic oscillations. It was pointed out that the
outflows, and not accretion disks, could be responsible for the Iron K$_\alpha$ lines
and the so-called reflection components [32]. A schematic representation of the detailed
picture as above is drawn in Fig. 2b. We shall use this picture below to analytically 
estimate the outflow rates.

A new concept which was found useful in identifying black holes by using spectral features
alone is the `bulk motion Comptonization'. Basically, as matter flows into a black hole
rapidly with almost the speed of light, photons scattering from them gain momentum and 
frequency is blue-shifted due to Doppler effect [32, 39].
The resulting spectra is of power-law and extends till about an MeV. 
The presence of this part of the spectra in soft states is widely accepted to be the only way to
identify a black hole most convincingly, whether galactic or extragalactic, since the
all absorbing property of the horizon is directly used in obtaining the spectral slope.

Black holes in X-ray binaries often show quiescence states. This could be the result of
very low accretion rates. A novel solution that was proposed by Das \& Chakrabarti [40]
(also see, [41-42]) is that outflows generated from the inflow could, in some
certain circumstances especially when the accretion rate is low, 
be so high that the disk may be almost evacuated. They proposed that 
such outflows could generate quiescence states of a black hole candidate. 

\section{Global Inflow-Outflow Solutions (GIOS)}

Outflows are common in many astrophysical systems which 
contain black holes and neutron stars. Difference between stellar
outflows and outflows from these systems is that the outflows in these
systems have to form out of the inflowing material only. We now present
a simple analytical approach by which the outflow
rate is computed out of the inflow rate. The ratio of these two rates is
found to be a function the compression ratio of the gas at the boundary
between CENBOL and the inflow.

The problem of mass outflow rate in the context of a black hole accretion
has been attacked quite recently [27, 43]. A simple approach
widely used in stellar astrophysics has been followed where the flow upto the
sonic point is assumed to be isothermal. This was possible due to the
novel understanding that around a black hole, the centrifugal
pressure supported dense matter could behave like a `boundary layer'. 
This CENBOL is hot, puffed up and very similar to 
the classical thick accretion disk, except that at the inner 
edge the matter is strongly advective. The thermal pressure is 
expected to drive matter out in between the centrifugal barrier 
and the funnel wall [31] as in Fig 2a. However, we use a simple model
(Fig. 2b) where both the inflow is axially symmetric and wedge shaped 
while the outflow is conical with solid angles $\Theta_{in}$ and $\Theta_{out}$ respectively.

CENBOL could form either with or without shocks as long as the 
angular momentum of the flow close to the compact object is roughly
constant as is the case in reality [22-24]. This region replaces [32, 44-45]
the so called `Compton cloud' in explaining hard and soft states of black hole. 
The oscillation of this region can explain the general properties of the 
quasi-periodic oscillation [37,38,46] from black holes and neutron stars. 
It is therefore of curiosity if this region plays any major 
role in formation of outflows.

Several authors have also mentioned of denser regions  closer to a black hole due to different physical effects.
Chang \& Ostriker [47] showed that pre-heating of the gas could  produce standing
shocks at a large distance (typically a few tens of thousands Schwarzschild
radii away).  Kazanas \& Ellison [48] mentioned 
that pressure due to pair plasma could produce standing shocks at smaller distances 
around a black hole as well. Our computation is insensitive to the actual 
mechanism by which the boundary layer is produced. All we require is that the gas should
be hot at the region where the compression takes place. Thus, since Comptonization
processes cool this region [32] for larger accretion rates (${\dot M} \gsim 0.1 
{\dot M_{Eddington}}$) our process is valid only for low-luminosity objects, 
consistent with current observations. Begelman \& Rees [12] talked about 
a so-called `cauldron' model of compact objects where jets were assumed to
emerge from a dense mixture of matter and radiation by boring a de-Laval nozzle as in 
Blandford \& Rees [9] model. The difference between this model and the 
present one is that very high accretion 
rate was required (${\dot M}_{in} \sim 1000 {\dot M}_E$)  there while we consider thermally driven outflows
out of accretion with smaller rates. Second, the size of the `cauldron' was thousands of
Schwarzschild radii (where gravity was so weak that channel has to have shape of
a de-Laval nozzle to produce transonicity), while we have a CENBOL of about $10 R_g$ (where the gravity
plays an active role in creating the transonic wind) in our mind. 
Third, in the present case, matter is assumed to pass through a sonic 
point using the pre-determined funnel where rotating pre-jet matter is 
accelerated [16] and not through a `bored nozzle'
even though symbolically a quasi-spherical CENBOL is considered for mathematical convenience.

Once the presence of the centrifugal pressure supported boundary layer (CENBOL) 
is accepted, the mechanism of the formation of the outflow becomes clearer. One basic criteria
is that the outflowing winds should have positive Bernoulli constant [22]. Just as photons
from the stellar surface deposit momentum on the outflowing wind and keeps the flow
roughly isothermal [49] at least upto the sonic point, one may assume 
that the outflowing wind close to the black hole is kept isothermal due to 
deposition of momentum from hard photons. 
In the case of the sun, it's luminosity is only $10^{-5}\ L_{Edd}$ and the typical mass outflow
rate from the solar surface is $10^{-14}M_\odot$ year$^{-1}$ [50]. Proportionately, for a
star with a Eddington luminosity, the outflow rate would be $10^{-9} M_\odot$ year$^{-1}$. This is
roughly half as the Eddington rate for a stellar mass star. Thus if the flow is 
compressed and heated at the centrifugal barrier around a black hole, it would also
radiate enough to keep the flow isothermal (at least up to the sonic point) if the efficiency
were exactly identical. Physically, both requirements may be equally
difficult to meet, but in reality with photons shining on outflows near a black hole with almost
$4\pi$ solid angle (from funnel wall) it is easier to maintain the isothermality in the slowly moving
(subsonic) region in the present context. Another reason is  
this: the process of momentum deposition on electrons is more efficient near a black hole.
The electron density $n_e$ falls off as $r^{-3/2}$ while the photon density $n_\gamma$
falls off as $r^{-2}$. Thus the ratio $n_e/n_\gamma \propto r^{1/2}$ increases with the size of the region.
Thus a compact object will have lesser number of electrons per photon and the momentum transfer is
more efficient. In a simpler minded way the physics is scale-invariant, though. In solar physics, it is
customary to chose a momentum deposition term which keeps the flow isothermal to be of the
form [51],
$$
F_r = \int_{R_s}^r D dr
$$
where, $D$ is the momentum deposition (localized around $r_p$) factor with a typical spatial dependence,
$$
D=D_0 e^{-\alpha (r/r_p-1)^2}
$$ 
Here, $D_0$, $\alpha$ are constants and $R_s$ is the location of the stellar surface.
Since $r$ and $r_p$ comes in ratio, exactly same physical consideration would be
applicable to black hole physics, with the same result {\it provided} $D_0$ is scaled with
luminosity (However, as we showed above, $D_0$ goes up for a compact object due to higher
solid angle.). However, as Chakrabarti \& Titarchuk [32] showed, high accretion rate (${\dot M} \gsim 0.3 {\dot M}_{Edd}$ )
will {\it reduce} the  temperature of the CENBOL catastrophically, and therefore our assumption of
isothermality of the outflow would severely breakdown at these high rates simply because cooler outflow
would have sonic point very far away from the black hole. It is to be noted that
in the context of stellar physics, it is shown [52] that the temperature
stratification is the outflowing wind has little effect on the mass loss rate. Effect of radiation momentum
deposition on the topology of the  outflows is separately discussed in Chattapadhyay [53].

\subsection{Derivation of the outflow rate using simple GIOS}

The accretion rate of the incoming accretion flow is given by,
$$
{\dot M}_{in} = \Theta_{in} \rho \vartheta r^2 .
\eqno{(1)}
$$
Here, $\Theta_{in}$ is the solid angle subtended by the inflow, $\rho$ and
$\vartheta$ are the density and velocity respectively, and $r$ is the
radial distance. For simplicity, we assume geometric units ($G=1=M_{BH}=c$;
$G$ is the gravitational constant, $M_{BH}$ is the mass of the central black hole,
and $c$ is the velocity of light) to measure all the quantities. 
In this unit, for a freely falling gas,
$$
\vartheta (r)= [\frac{1-\Gamma}{r}]^{1/2}
\eqno{(2)}
$$
and
$$
\rho(r) = \frac {{\dot M}_{in}}{\Theta_{in}}(1-\Gamma)^{-1/2} r^{-3/2}
\eqno{(3)}
$$
Here, $\Gamma/r^2$ (with $\Gamma$ assumed to be a 
constant) is the outward force due to radiation. 

We assume that the boundary of the denser cloud is at $r=r_s$
(typically a few Schwarzschild radii, see, Chakrabarti [24])
where the inflow gas is compressed. The compression could be
abrupt due to standing shock or gradual as in a shock-free flow
with angular momentum. This details are irrelevant. At this barrier, then 
$$
\rho_+(r_s) = R \rho_- (r_s) 
\eqno{(4a)}
$$
and 
$$
\vartheta_+(r_s) = R^{-1} \vartheta_- (r_s) 
\eqno{(4b)}
$$
where, $R$ is the compression ratio.  Exact value of the compression ratio
is a function of the flow parameters, such as the specific energy and the
angular momentum (e.g., [22-24])
Here, the subscripts $-$ and $+$ denote the pre-shock and post-shock 
quantities respectively. At the shock surface, the total pressure 
(thermal pressure plus ram pressure) is balanced.
$$
P_- (r_s) + \rho_- (r_s) \vartheta_-^2 (r_s)
= P_+ (r_s) + \rho_+ (r_s) \vartheta_+^2 (r_s).
\eqno{(5)}
$$
Assuming that the thermal pressure of the pre-shock incoming flow is 
negligible compared to the ram pressure, using eqs. 4(a-b) we find,
$$
P_+(r_s) = \frac{R-1}{R} \rho_-(r_s) \vartheta_-^2 (r_s).
\eqno{(6)}
$$
The isothermal sound speed in the post-shock region is then,
$$
C_s^2= \frac{P_+}{\rho_+}=\frac{(R-1)(1-\Gamma)}{R^2}\frac{1}{r_s}
=\frac{(1-\Gamma)}{f_0 r_s}
\eqno{(7)}
$$
where, $f_0=R^2/(R-1)$. 
An outflow which is generated from this dense region with very low flow
velocity along the axis is necessarily subsonic in this region,
however, at a large distance, the outflow velocity is expected to be
much higher compared to the sound speed, and therefore the flow must be
supersonic. In the subsonic region of the outflow, the pressure and density
are expected to be almost constant and thus it is customary to 
assume isothermality condition up to the sonic point [49].
With isothermality assumption or a given temperature
distribution ($T \propto r^{-\beta}$ with $\beta$ a constant; see eq. [22] below) the result 
is derivable in analytical form. The sonic point conditions are obtained 
from the radial momentum equation, 
$$
\vartheta \frac{d\vartheta}{dr} + \frac{1}{\rho}\frac{dP}{dr} 
+\frac{1-\Gamma}{r^2} = 0 
\eqno{(8)}
$$
and the continuity equation
$$
\frac{1}{r^2}\frac{d (\rho \vartheta r^2)}{dr} =0
\eqno{(9)}
$$
in the usual way, i.e., by eliminating $d\rho/dr$,
$$
\frac{d\vartheta}{dr}= \frac{N}{D}
\eqno{(10)}
$$
where
$$
N=\frac{2 C_s^2}{r} - \frac{1-\Gamma}{r^2}
$$
and
$$
D=\vartheta - \frac{C_s^2}{\vartheta}
$$
and putting $N=0$ and $D=0$ conditions. These conditions
yield, at the sonic point $r=r_c$, for an isothermal flow,
$$
\vartheta (r_c) = C_s .
\eqno{(11a)}
$$
and
$$
r_c = \frac{1-\Gamma}{2 C_s^2}=\frac {f_0 r_s}{2}
\eqno{(11b)}
$$
where, we have utilized eq. (7) to substitute for $C_s$. 

Since the sonic point of a hotter outflow
is located closer to the black hole, clearly, the condition of isothermality
is best maintained if the temperature is high enough. However if the temperature
is too high, so that $r_c <r_s$, one has to solve this case
more carefully, using considerations of Fig. 2a, rather than of Fig. 2b.

The constancy of the integral of the radial momentum equation 
(eq. 8) in an isothermal flow gives: 
$$
C_s^2 ln \ \rho_+ -\frac{1-\Gamma}{r_s} =
\frac{1}{2}C_s^2 + C_s^2 ln \ \rho_c -\frac{1-\Gamma}{r_c}
\eqno{(12)}
$$
where, we have ignored the initial value of the outflowing 
radial velocity $\vartheta (r_s)$ at the dense region boundary ($r=r_s$), 
and also used eq. (11a). We have also put $\rho(r_c)=\rho_c$ 
and $\rho(r_s) = \rho_+$. Upon simplification, we obtain,
$$
\rho_c =\rho_+  exp (-f)
\eqno{(13)}
$$
where,
$$
f= f_0 - \frac{3}{2}
$$
Thus, the outflow rate is given by,
$$
{\dot M}_{out} = \Theta_{out} \rho_c \vartheta_c r_c^2 
\eqno{(14)}
$$
where, $\Theta_{out}$ is the solid angle subtended by the outflowing cone. 
Upon substitution, one obtains,
$$
\frac{{\dot M}_{out}} {{\dot M}_{in}} = R_{\dot m}
=\frac{\Theta_{out}}{\Theta_{in}} \frac{R}{4} f_0^{3/2} exp \ (-f)
\eqno{(15)}
$$
which, explicitly depends only on the compression ratio:
$$
\frac{{\dot M}_{out}}{{\dot M}_{in}} =R_{\dot m}=
\frac{\Theta_{out}}{\Theta_{in}}\frac{R}{4} 
[\frac{R^2}{R-1}]^{3/2} exp  (\frac{3}{2} - \frac{R^2}{R-1})
\eqno{(16)}
$$
apart from the geometric factors. Notice that this simple result 
does not depend on the location of the sonic points or the
the size of the dense cloud or the outward radiation 
force constant $\Gamma$. This is because the Newtonian potential 
was used throughout and the radiation force was also assumed 
to be very simple minded ($\Gamma/r^2$). Also, 
effects of centrifugal force  was ignored.  Similarly, the ratio
is independent of the mass accretion rate which should be valid only for
low luminosity objects. For high luminosity flows, Comptonization would
cool the dense region completely [32] and the mass loss will be negligible.
Pair plasma supported quasi-spherical shocks forms for low luminosity as well [48].
In reality there would be a dependence on these 
quantities when full general relativistic considerations of the rotating flows are 
made. Exact and detailed computations using both the transonic inflow
and outflow (where the compression ratio $R$ is also computed self-consistently)
are presented elsewhere [40].

Figure 3 contains the basic results. The solid curve shows the
ratio $R_{\dot m}$ as a function of the compression ratio $R$ (plotted from $1$ to $7$),
while the dashed curve shows the same quantity as a function of the
polytropic constant $n=(\gamma-1)^{-1}$ (drawn from $n=3/2$ to $3$), $\gamma$ being the adiabatic
index. The solid curve is drawn for any generic compression ratio and the dashed curve is 
drawn assuming the strong shock limit only: 
$R=(\gamma+1)/(\gamma-1)=2n+1$. In both the curves, $\Theta_{out} \sim
\Theta_{in}$ has been assumed for simplicity. Note that if the compression 
does not take place (namely, if the denser region does not form), then
there is no outflow in this model. Indeed for, $R=1$, the ratio 
$R_{\dot m}$ is zero as expected. Thus the driving force of the outflow 
is primarily coming from the hot and compressed region. The basic form 
is found to agree with results obtained from rigorous calculation with 
transonic inflow and outflow.    

In a relativistic inflow or for a radiation dominated inflow, $n=3$ and 
$\gamma=4/3$. In the strong shock limit, the compression ratio is $R=7$ 
and the ratio of inflow and outflow rates becomes,
$$
R_{\dot m}=0.052 \ \frac{\Theta_{out}}{\Theta_{in}}.
\eqno{(17a)}
$$
For the inflow of a mono-atomic ionized gas $n=3/2$ and $\gamma=5/3$. 
The compression ratio is $R=4$, and the ratio in this case becomes,
$$
R_{\dot m}=0.266 \ \frac{\Theta_{out}}{\Theta_{in}}.
\eqno{(17b)}
$$
Since $f_0$ is smaller for $\gamma=5/3$ case, the density at the
sonic point in the outflow is much higher 
(due to exponential dependence of density on $f_0$, see, eq. 7) 
which causes the higher outflow rate, even when the actual jump in density
in the postshock region, the location of the
sonic point and the velocity of the flow at the sonic point are much lower.
It is to be noted that generally for $\gamma >1.5$ shocks are not
expected [23], but the centrifugal barrier supported dense region
would still exist. As is clear, the entire behavior of the outflow
depends only on the compression ratio, $R$ and the collimating
property of the outflow $\Theta_{out}/\Theta_{in}$.

Outflows are usually concentrated near the axis, while the inflow is near
the equatorial plane. Assuming a half angle of $10^o$ in each case, 
we obtain,
$$
\Theta_{in}= \frac {2 \pi^2}{9}; \ \ \ \ \ \Theta_{out}= \frac {\pi^3}{162}
$$
and 
$$
\frac{\Theta_{out}}{\Theta_{in}} =\frac{\pi}{36} .
\eqno{(18)}
$$
The ratios of the rates for $\gamma=4/3$ and $\gamma=5/3$ are then
$$
R_{\dot m}=0.0045
\eqno{(19a)}
$$
and 
$$
R_{\dot m}= 0.023
\eqno{(19b)}
$$
respectively. Thus, in quasi-spherical systems, 
in the case of strong shock limit, the outflow rate is at the most a couple 
of percent of the inflow. If this assumption is dropped, then for a cold inflow,
the rate could be higher by about fifty percent (see, Fig. 3).

It is to be noted that the above expression for the outflow rate is strictly
valid if the flow could be kept isothermal at least up to the sonic point. In the
event this assumption is dropped the expression for the outflow rate becomes
dependent on several parameters. As an example, we consider a 
polytropic outflow of  same index $\gamma$ but of a 
different entropy function $K$ (We assume the equation of state to be $P=K\rho^\gamma$, with
$\gamma\neq 1$). The expression (11b) would be replaced by,
$$
r_c=\frac{f_0r_s}{2\gamma}
\eqno{(20)}
$$
and eq. (12) would be replaced by,
$$
n a_+^2 - \frac{1-\Gamma}{r_s}=(\frac{1}{2} + n) a_s^2 - \frac{1-\Gamma}{r_c}
\eqno{(21)}
$$
where $n=1/(\gamma-1)$ is the polytropic constant of the flow
and $a_+=(\gamma P_+/\rho_+)^{1/2}$ and $a_c=(\gamma P_c/\rho_c)^{1/2}$ 
are the adiabatic sound speeds at the starting point and 
the sonic point of the outflow. It is easily shown that
a power law temperature fall off of the outflow ($T\propto r^{-\beta}$) would yield
$$
R_{\dot m}= \frac{\Theta_{out}}{\Theta_{in}} (\frac{K_i}{K_o})^n 
(\frac{f_0}{2\gamma})^{\frac{3}{2}-\beta} ,
\eqno{(22)}
$$
where, $K_i$ and $K_o$ are the entropy functions of the inflow and the outflow. This derivation is
strictly valid for a non-isothermal flow. Since $K_i<K_o$, $n>3/2$ and $f_0$,
for ${\Theta_{out}} \sim {\Theta_{in}}$, $R_{\dot m} <<1$ is guaranteed provided $\beta >\frac{3}{2}$,
i.e., if the temperature falls for sufficiently rapidly. For an isothermal flow $\beta=0$ and the rate 
tends to be higher. Note that since $n\sim \infty$ in this case, any small
jump in entropy due to compression will off-balance the the effect of $f_0^{-3/2}$ factor.
Thus $R_{\dot m}$ remains smaller than unity. The first factor decreases with entropy
jump while the second factor increases with the compression ratio ($R$) when $\beta<3/2$.
Thus the solution is still expected to be similar to what is shown in Fig. 3.

\subsection{Results from Rigorously obtained GIOS}

When inflow and outflow solutions are obtained more rigorously, i.e., actually making the
flow pass through separate sonic surfaces, one has to include the effect of mass-loss
on the Rankine-Hugoniot condition  at the boundary between CENBOL and the accretion flow [40]. Accordingly, we use,
$$
{\dot M}_+ = (1-R_{\dot m}) {\dot M}_-
\eqno{(23)}
$$
where, the subscripts $+$ and $-$ denote the pre- and post-shock values respectively. Since due to the
loss of matter in the post-shock region, the post-shock pressure goes down, the shock recedes
backward for the same value of incoming energy, angular momentum \& polytropic index. The combination of
three changes, namely, the increase in the cross-sectional area of the outflow and the
launching velocity of the outflow and the decrease in the post-shock density decides whether the
net outflow rate would increased or decreased than from the case when the
exact Rankine-Hugoniot relation was used.

Fig. 4 shows a typical global inflow-outflow solution (GIOS) where actual transonic
problem was solved. The input parameters
are ${\cal E}=0.0005$, ${\lambda=1.75}$ and $\gamma=4/3$ corresponding to
relativistic inflow. The solid curve with an arrow represents
the pre-shock region of the inflow and the long-dashed curve represents the post-shock
inflow which enters the black hole after passing through the inner sonic point (I).
The solid vertical line at $X_{s3}$ (the leftmost vertical transition; the notation 
of the shock location is from [22]) with
double arrow represents the shock transition obtained
with exact Rankine-Hugoniot condition (i.e., with no mass loss).
The actual shock location obtained with modified Rankine-Hugoniot condition
(eq. 23) is farther out from the original location $X_{s3}$.
Three vertical lines connected with the corresponding dotted curves represent
three outflow solutions for the parameters $\gamma_{o}=1.3$
(top), $1.15$ (middle) and $1.05$ (bottom). The outflow
branches shown pass through the corresponding sonic points. It is
evident from the figure that the outflow  moves along solution curves which are completely different from that
of the `wind solution' of the inflow which passes through the outer sonic point `O'.
The mass loss ratio $R_{\dot m}$ in these cases are $0.256$, $0.159$ and $0.085$ respectively. 
This figure is taken from [40].
Fig. 3 of that work shows the general behaviour of the variation of $R_{\dot M}$ with compression ratio of the shock.
This agrees well with what is obtained analytically (Fig. 3 of this paper).

Das \& Chakrabarti [40], for the first time, pointed out that in certain region
of the parameter space, the mass loss could be so much that  disk would be almost
evacuated and conjectured that inactive phases of a black hole (such as the present day Sgr A* at our galactic center) 
may be formed when such profuse mass loss takes place.

It is to be noted that although the existence of outflows are well known,
their rates are not. The only definite candidate whose outflow rate is known with
any certainty is probably SS433 whose mass outflow rate was estimated to be
${\dot M}_{out} \gsim 1.6 \times 10^{-6}  f^{-1} n_{13}^{-1} D_5^2 M_{\odot} $ yr$^{-1}$
[54], where $f$ is the volume filling factor, $n_{13}$
is the electron density $n_e$ in units of $10^{13}$ cm$^{-3}$, $D_5$
is the distance of SS433 in units of $5$kpc. Considering a central
black hole of mass $10M_{\odot}$, the Eddington rate is ${\dot M}_{Ed} \sim
0.2 \times 10^{-7} M_{\odot} $ yr$^{-1}$ and assuming an efficiency
of conversion of rest mass into gravitational energy $\eta \sim 0.06$, the
critical rate would be roughly ${\dot M}_{crit} = {\dot M}_{Ed} / \eta \sim
3.2 \times 10^{-7} M_{\odot} $ yr$^{-1}$. Thus, in order to produce the outflow rate
mentioned above even with our highest possible estimated $R_{\dot m}\sim 0.4$ (see,
Fig. 2), one must have ${\dot M}_{in} \sim 12.5 {\dot M}_{crit}$ which is very high
indeed. One possible reason why the above rate might have been over-estimated
would be that below $10^{12}$cm from the central mass [54], $n_{13} >>1 $
because of the existence of the dense region at the base of the outflow.

In numerical simulations the ratio of the outflow and inflow has been computed
in several occasions [19,29]. Eggum et al. [19]  found the ratio to be $R_{\dot m} \sim 0.004$ for a
radiation pressure dominated flow. This is generally comparable with what we found
above (eq. 19a). In Molteni et al. [29] the centrifugally driven outflowing wind
generated a ratio of $R_{\dot m}\sim 0.1$. Here, the angular momentum was present
in both inflow as well as outflow, and the shock was not very strong. Thus,
the result is again comparable with what we find here.

\subsection{Spectral Softening in Presence of Profuse Mass Loss}

Creation of winds out of accretion disk has an interesting consequence.
As more fraction of matter is expelled, it becomes 
easier for the soft photons from the Keplerian disk (see Fig. 2a) surrounding the advective 
region to cool this region due to Comptonization (see, Chakrabarti \& Titarchuk 
for details). In other words, the presence of winds
would {\it soften} the spectra of the power-law component for the {\it same}
multicolour blackbody component. Such an observation
would point to profuse mass loss from the hot advective region. 

The model we use here is the two component accretion flow which has 
a CENBOL close to the hole. We assume
a weakly viscous flow of constant specific angular momentum $\lambda=1.75$
(in units of $2GM/c$) for the sake of concreteness. The Keplerian
component close to the equatorial plane has a low accretion rate
(${\dot M}_{in} \sim 0.05-0.3$ in units of the Eddington rate) 
and the sub-Keplerian halo surrounding it has a higher rate 
(${\dot M}_h \sim 1$ in units of Eddington rate). Before the 
accreting matter hits the inner advective region, both the rates are constant,
but as Das \& Chakrabarti [40] has shown, winds, produced from CENBOL
will deplete the disk matter at the rate determined by the temperature of the
CENBOL, when other parameters, such as the specific angular momentum and specific
energy are kept fixed. 

Figures 5 and 6 show the outcome of our calculation of the spectra for three
different accretion rate of the Keplerian component ${\dot M}_{in}$ [43]. The mass
of the central black hole is chosen to be $M=10M_\odot$. The 
size of the CENBOL is assumed to be $10r_g$ (where $r_g$ is the
Schwarzschild radius), a typical location for the sub-Keplerian flow
of average specific angular momentum $\lambda=1.75$ and specific 
energy ${\cal E}=0.003$. Following Das \& Chakrabarti [40], 
we first compute the mass outflow rate
from the advective region. The long dashed curve in Fig. 5 shows the 
variation of the percentage of mass loss (vertical axis on the right)
as a function of the inflow accretion rate. The dotted curve and the
solid curve denote the variation  of the energy spectral index $\alpha$
($F_\nu \propto \nu^{-\alpha}$) with and without winds taken into account.
Note that  the spectra is overall softened ($\alpha$ increased)
when winds are present. For higher Keplerian rates, the mass loss through winds is 
negligible and therefore there is virtually no change in the spectral
index. For lower inflow rates, on the other hand, mass loss rate is more than twenty
percent. It is easier to Comptonize the depleted matter by the same number
of incoming soft photons and therefore the spectra is softened.

In Fig. 6, we show the resulting spectral change. As in Fig. 5, solid
curves represent solutions without winds and the dotted curves
represent solutions with winds. Solid curves are drawn for ${\dot M}_{in}=0.3$ 
(uppermost at the bump), $0.15$ (middle at the bump) and $0.07$ (lowermost at the bump)
respectively. For ${\dot M}_{in}=0.3$ both curves are identical.
Note the crossing of the solid curves at around $10^{18.6}Hz$
($15$ keV) when winds are absent. This is regularly observed in black hole
candidates. If this is shifted to higher energies, the presence of 
winds may be indicated.

Strong winds are suspected to be present in Sgr $A^*$ at our Galactic Center
(see, Genzel et al. [55] for a review, and Eckart \& Genzel [56]). Chakrabarti [9]
suggested that the inflow could be of almost constant energy transonic
flow, so that the emission is inefficient. However, from global inflow-outflow
solutions [GIOS], Das \& Chakrabarti [40] showed that when the inflow rate
itself is low (as is the case for Sgr A$^*$; $\sim 10^{-3}$ to $10^{-4} {\dot M}_{Eddington}$) 
the mass outflow rate is very high, almost to the point of evacuating the disk.
This prompted them to speculate that spectral properties of our
Galactic Center could be explained by inclusion of winds. This 
will be done in near future. Not only our Galactic Center, the consideration
should be valid for all the black hole candidates (e.g., V404 Cyg) which are seen
in quiescence. 

Chakrabarti \& Titarchuk [32] suggested that the iron K$_\alpha$ line as well 
as the so called `reflection component' could be due to outflows
off the advective region. Combined with the present work, we may conclude 
that simultaneous enhancement of the `reflection component' and/or 
iron K$_\alpha$ line intensity with the softening of the spectra 
in hard X-rays would be a sure signature of the presence of 
significant winds in the advective region of the disk.

\newpage

\centerline{\bf References}

\noindent 1. Shakura, N.I. \& Sunyaev, R.A. 1973, Black holes in binary systems: Observational appearance, Astr. Ap., {\bf 24}, 337-355\\
2. Longair, M.S., Ryle, M. \& Scheuer, P.A.G., 1973,Models of extended radio sources, Mon. Not. R. Astron. Soc. {\bf 164}, 243-250\\
3. Maraschi, L., Reina C., \& Treves, A., 1976, 
The effect of radiation pressure on accretion disks around black holes, Astrophys. J. {\bf 206}, 295-300\\
4. Paczy\'nski B. \& Wiita, P.J., 1980, Thick accretion disks and supercritical luminosities, Astron. Ap. {\bf 88}, 23-31\\
5. Liang E.P.T. \& Thompson, K.A. 1980, Transonic disk accretion onto black holes, Astrophys. J., {\bf 240}, 271-274\\
6. Paczy\'nski B. \& Bisnovatyi-Kogan, G. 1981, Acta Astron. {\bf 31}, 283-293\\
7. Muchotrzeb B., \& Paczy\'nski, B. 1982, Transonic accretion flow in a thin disk around a black hole, Acta Astron. {\bf 32}, 1-11\\
8. Chakrabarti, S. K. 1996, in Accretion Processes on Black Holes, Physics Reports, {\bf 266}, No. 5 \& 6, p. 229-390\\
9. Blandford, R.D. \& Rees, M.J. 1974, A 'twin-exhaust' model for double radio sources, Mon. Not. R. Astro. Soc., {\bf 169}, 395-415\\
10. Znajek, R.L., 1978, Charged current loops around Kerr holes,  Mon. Not. R. Astron. Soc., {\bf 182}, 639-646\\
11. Blandford R.D. \& Payne, D.G. 1981, Hydromagnetic flows from accretion discs and the production of radio jets,
 Mon. Not. R. Astron. Soc. {\bf 194}, 883-903\\
12. Begelman, M.C. \& Rees, M.J., 1984, The cauldron at the core of SS 433, Mon. Not. R. Astro. Soc., {\bf 206}, 209-220\\
13. K\"onigl, A. 1989, Self-similar models of magnetized accretion disks, Astrophys. J., {\bf 342}, 208-223 \\
14. Chakrabarti, S.K. \& Bhaskaran, P. 1992, Mon. Not. R. Astron. Soc. {\bf 255}, 255-260 \\
15. Contopoulos, J. 1995, Force-free Self-similar 
Magnetically Driven Relativistic Jets Astrophys. J., {\bf 446}, 67-74 \\
16. Chakrabarti, S.K. 1984 in Active Galactic Nuclei, ed. J. Dyson, (Manchester University Press), 346-350\\
17. Chakrabarti, S.K., 1985 Astrophys. J. {\bf 288}, 7-13\\
18. Hawley, J.W., Wilson, J. \& Smarr, L.  1984, A numerical study of nonspherical black hole accretion: I-
Equations and test problems, Astrophys. J., {\bf 277}, 296-311\\
19. Eggum, G. E., Coroniti, F. V., Katz, J. I. 1985, Jet production in super-Eddington accretion disks,
 Astrophys. J., {\bf 298}, L41-L45\\
20. Bondi, H. 1952, On Spherically Symmetrical Accretion, Mon. Not. R. Astron. Soc. {\bf 112}, 195-199\\
21. Parker, E.N., 1979, Cosmical Magnetic Fields, (Clarendon Press: Oxford)\\
22. Chakrabarti, S. K. 1989, Astrophys. J., {\bf 347}, 365-372 \\
23. Chakrabarti, S. K. 1990, Theory of Transonic Astrophysical Flows (Singapore: World Sci.) \\
24. Chakrabarti, S. K. 1996, Astrophys. J., {\bf 471}, 237-247 \\
25. Chakrabarti, S.K. 1990 Mon. Not. R. Astron. Soc. {\bf 243}, 610-619\\
26. Chakrabarti, S. K. 1996, Astrophys. J., {\bf 464}, 664-683 \\
27. Chakrabarti, S.K., 1998 in `Observational Evidence for Black Holes in the Universe', 
ed. S.K. Chakrabarti (Kluwer Academic Publishers: Dordrecht), 19-48 astro-ph/9807104\\
28. Chakrabarti, S.K. \& Molteni D. 1993, Astrophys. J., {\bf 417}, 671-676\\
29. Molteni, D., Lanzafame, G. \& Chakrabarti, S. K. 1994, Simulation of thick accretion disks with standing shocks by
smoothed particle hydrodynamics, Astrophys. J., {\bf 425}, 161-170\\
30. Chakrabarti S.K. \& Molteni, D. 1995, Mon. Not. R. Astron. Soc. {\bf 272}, 80-88\\
31. Molteni, D., Ryu., D. \& Chakrabarti 1996, Numerical Simulations of Standing Shocks in Accretion
Flows around Black Holes: A Comparative Study, Astrophys. J., {\bf 470}, 460 \\
32. Chakrabarti, S. K., \& Titarchuk, L.G. 1995, Astrophys. J., {\bf 455}, 623-639 \\
33. Zhang, S.N., Cui, W., Harmon, B.A., Paciesas, W.S., Remillard R.E., \& van Paradijs J. 1997, The 1996 Soft State Transition of Cygnus X-1,
 Astrophys. J., {\bf 477}, L95-L100\\
34. Gilfanov, M., Churazov, E. \&  Sunyaev, R.A. 1997, Spectral and Temporal variations of the X-ray emission
from black hole and neutron star binaries, in {\it Accretion Disks -- New Aspects},
Eds. E. Meyer-Hofmeister \& H. Spruit, Springer (Heidelberg) \\
35. Sunyaev, R.A., et al., 1994, Observations Of X-Ray Novae In Vela 1993 Ophiuchus 1993 And Perseus 1992 Using
The Instruments Of The MIR / KVANT Module,  Astron. Lett., {\bf 20}, 777-784\\
36. Ebisawa, K. et al. 1994, Spectral evolution of the bright X-ray nova GS 1124-68
(Nova Muscae 1991) observed with Ginga, P.A.S.J., {\bf 46}, 375-394\\
37. Molteni, D., Sponholz, H. \& Chakrabarti, S. K. 1996, X-rays from Shock Waves In Accretion Flows Around
Black Holes, Astrophys. J., {\bf 457}, 805-812\\
38. Ryu, D., Chakrabarti, S. K., \& Molteni, D. 1997, Zero-Energy 
Rotating Accretion Flows near a Black Hole, Astrophys. J., {\bf 474}, 378-388\\
39. Titarchuk, L.G., Mastichiadis, A., \& Kylafis, N. G. 1997, X-Ray Spectral Formation in a Converging Fluid 
Flow: Spherical Accretion into Black Holes, Astrophys. J., {\bf 487}, 834-850 \\
40. Das T. \& Chakrabarti S.K. 1997, Astrophys. J. (submitted) astro-ph/9809109\\
41. Das, T.K. 1998a, `Observational Evidence for Black Holes in the Universe' 
ed. S.K. Chakrabarti (Kluwer Academic: Holland) p. 113-122 astro-ph/9807105\\
42. Das, T.K. 1998b, (this volume)\\
43. Chakrabarti, S. K. 1998, Indian Journal of Physics, {\bf 72B}, 565-569, astro-ph/9810412\\
44. Chakrabarti, S. K., Titarchuk, L.G., Kazanas, D. \& Ebisawa, K. 1996, A \& A Supp. Ser., {\bf 120}, 163-166 \\
45. Chakrabarti S.K., 1997 Astrophys. J., {\bf 484}, 313-322 \\
46. Paul, B., Agrawal, P.C., Rao, A.R., et al., 1998, Quasi-regular X-Ray Bursts from GRS 1915+105 Observed
with the IXAE: Possible Evidence for Matter Disappearing into
the Event Horizon of the Black Hole, Astrophys. J., {\bf 492}, L63-L67\\
47. Chang, K.M. \& Ostriker, J.P. 1985, Standing shocks in accretion flows onto black holes, Astrophys. J., {\bf 288}, 428-437\\
48. Kazanas, D. \& Ellison, D. C. 1986, The central engine of quasars and active galactic nuclei
Hadronic interactions of shock-accelerated relativistic
protons, Astrophys. J., {\bf 304}, 178-187\\
49. Tarafdar, S.P. 1988, A unified formula for mass-loss rate of O to M stars,
Astrophys. J., {\bf 331}, 932-938\\
50. Priest, E.R., 1982, Solar Magnetohydrodynamics, (Dordrecht, Holland)\\
51. Kopp, R.A. \& Holzer, T.E., 1976, Dynamics of 
coronal hole regions. I - Steady polytropic flows with multiple critical points, 
Sol. Phys., {\bf 49}, 43-56\\
52. Pauldrach, A., Puls, J., and Kudritzki, R.P. 1986, Radiation-driven winds of hot luminous stars - Improvements of
the theory and first results, Astr. Ap., {\bf 164}, 86-100\\
53. Chattapadhyay, I 1998, (this volume)\\
54. Watson, M.G., Stewart, G. C., Brinkann, W., \& King, A. R. 1986, 
Doppler-shifted X-ray line emission from SS433, Mon. Not. R. Astro. Soc., {\bf 222}, 261-271\\
55. Genzel, R., Hollenbach D. \& Townes, C.H., 1994, The nucleus of our Galaxy. Rep. Prog. Phys.  {\bf 57}, 417-479\\
56. Eckart A. \& Genzel, R. 1997, Stellar proper 
motions in the central 0.1 pc of the Galaxy, Mon. Not. R. Astro. Soc., {\bf 284}, 576-598\\
\newpage

\begin{center}
{\bf FIGURE CAPTIONS}
\end{center}
\begin{description}

\item[Fig.~1]: Classification of the parameter space (central box)
in the energy-angular momentum plane in terms of various topology of the
black hole accretion. Eight surrounding boxes show the solutions from
each of the independent regions of the parameter space. Each small
box shows Mach number $M$ against the logarithmic radial distance $r$
(measured in units of $GM_{BH}/c^2$)
Contours are of constant entropy accretion rate
${\dot{\cal M}}$. Similar classification is possible for all adiabatic index
$\gamma <1.5$. For $\gamma >1.5$, only the inner sonic point is possible
other than an unphysical `O' type point (Chakrabarti 1996c).

\item[Fig.~2a]: Cartoon diagram of a very general inflow-outflow configuration
of non-magnetized matter around a compact object. Keplerian and sub-Keplerian
matter accretes along the equatorial plane. Centrifugally and thermally driven
outflows preferentially emerge between the centrifugal barrier
and the funnel wall. 

\item[Fig.~2b] Schematic diagram of inflow and outflow around a compact object. 
Hot, dense region around
the object either due to centrifugal barrier or due to plasma pressure effect
or pre-heating, acts like a `stellar surface' from which the outflowing wind is developed.

\item[Fig.~3] Ratio ${\dot R}_{\dot m}$ of the outflow rate and the inflow rate
as a function of the compression ratio of the gas at the dense region boundary
(solid curve). Also shown in dashed  curve is its variation with the polytropic
constant $n$ in the strong shock limit. Solid angles subtended by the inflow
and the outflow are assumed to be comparable (Chakrabarti 1998a).

\item[Fig.~4]: Variation of the percentage of mass loss (long dashed curve and right axis) 
and the energy spectral index $\alpha$ ($F_\nu \propto \alpha^{-\alpha}$) (solid and dotted curves and
left axis) with the accretion rate (in units of Eddington rate) of the Keplerian component.
Solid curve is drawn when winds are neglected from the advective region and dotted curve includes
effect of winds. Overall spectra is softened when the inflow rate is reduced (Chakrabarti
1998b).

\item[Fig.~5]: Spectra of emitted radiation from the accretion disk with (dotted) and without (solid)
effects of winds.  Hard X-ray component is softened while keeping soft  X-ray bump unchanged (Chakrabarti
1998b).

\item[Fig.~6]: Few typical solutions which combine accretion and outflow. Input parameters
are ${\cal E}=0.0005$, ${\lambda=1.75}$ and $\gamma=4/3$.
Solid curve with an incoming arrow represents
the pre-shock region of the inflow and the dashed curve with an incoming arrow
represents post-shock inflow which enters the black hole after passing through the inner sonic point
(I). Dotted curves are the outflows for various $\gamma_o$ (marked). Open circles are sonic points
of the outflowing winds and the crossing point `O' is the outer sonic point of the inflow. The
leftmost shock transition ($X_{s3}$) is obtained from unmodified Rankine-Hugoniot condition,
while  the other transitions are obtained when the mass-outflow is taken into account.

\end{description}

\clearpage

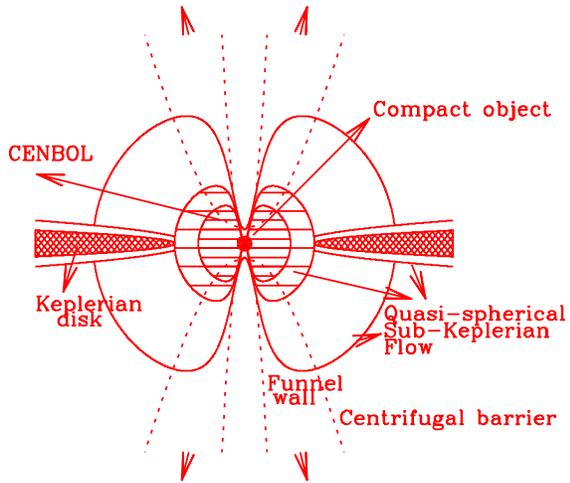
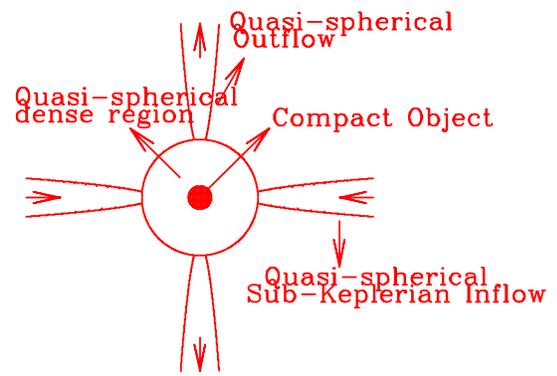
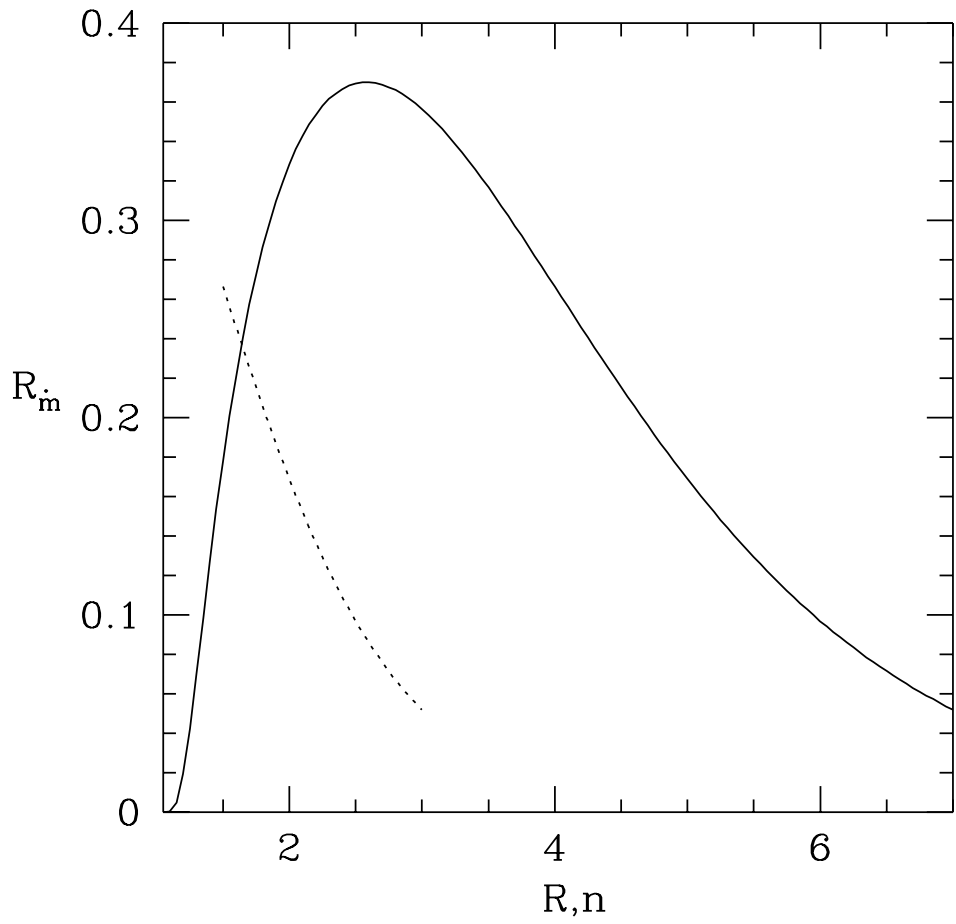
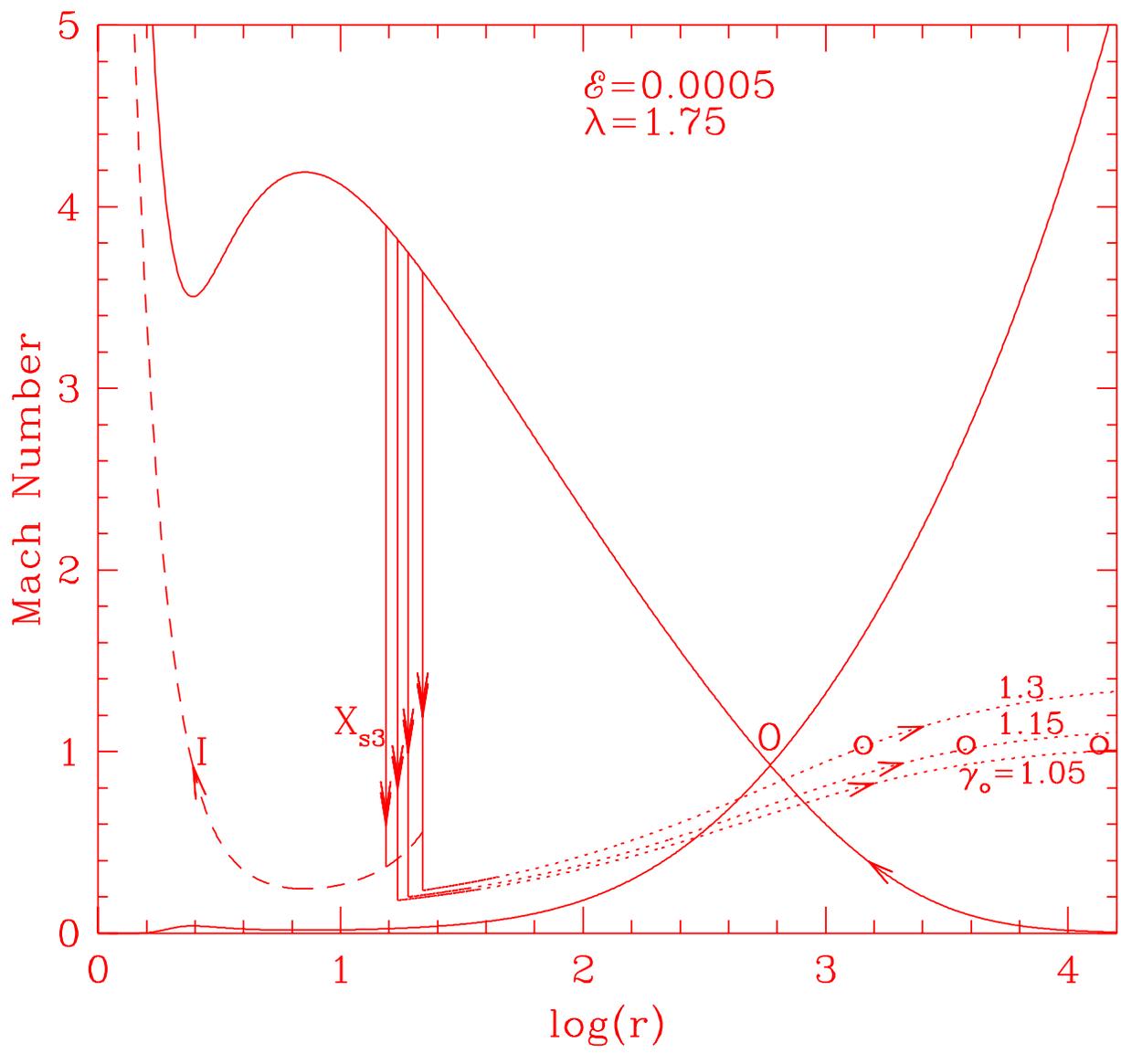
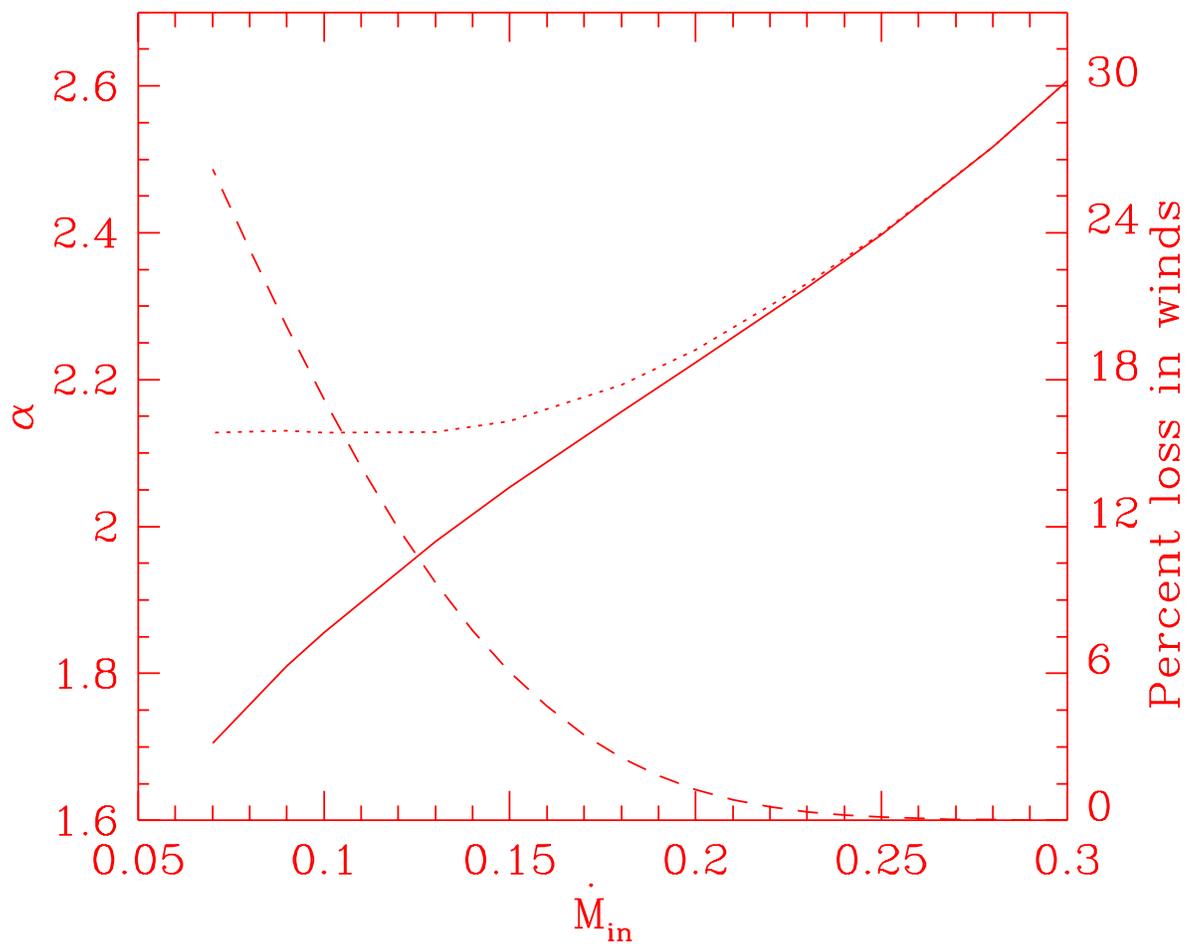
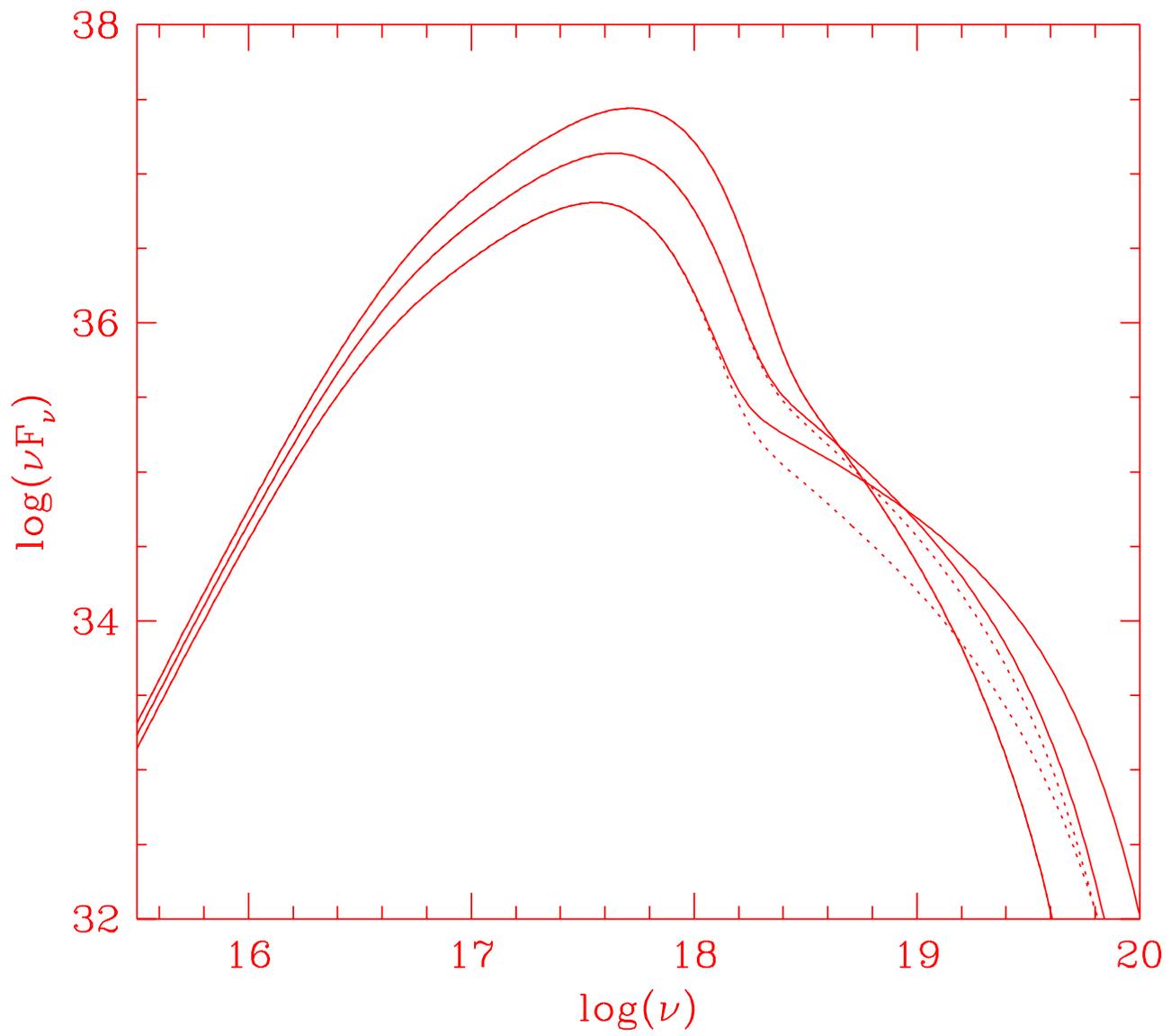
\end{document}